\documentstyle[preprint,prd,aps,epsf]{revtex}

\begin{document}
\draft
\title{\bf
$\bf{\triangle}$-N  ELECTROMAGNETIC TRANSITION}
\author{Mushtaq Loan\footnote{E-mail : mushe@newt.phys.unsw.edu.au}}  
\address{School of Physics, University of New South Wales, Sydney, Australia, 2052}
\date{January 25, 1999}
\maketitle
\begin{abstract}
 The EM ratio for a free $\bf{\triangle}$
electromagnetic transition is discussed within the frame work of 
nonrelativistic approach. Such an approach gives a good account of
data for a free $\bf{\triangle}$ but is less important for an intrinsically
relativistic nuclear many body problem.
\end{abstract}
\vspace{10mm}
\hspace{1.70cm} PACS number(s): 14.20.G\\
\begin{center}
\hspace{-3.65cm} Key words: Delta isobar, EM transition, EM ratio
\end{center}
\section{Introduction}
The free nucleon-nucleon(NN) interaction constitutes the basis of the 
microscopic approach to the description of the properties of the nuclear
system. Apart from the NN potential problem,
the conventional nuclear theory, in which nucleus is considered
as an assembly of bound nucleons only,  however, fails badly 
to account for the nuclear processes like photo
 and electro-disintegration of deutron and in charge and magnetic form 
 factors of three body nuclei, $\mbox{H}^{3}$ and $\mbox{He}^{3}$[1,2]
 in particular at higher momentum transfer. 
Thus it becomes evident that the only nucleon picture does not fully
simulate all physical effects of the hadronic fields even if hadronic fields
are replaced by the effective potential. The reason for this is that the
hadronic field is charged and therefore is subject to the electromagnetic
interaction, leading to the presence of meson exchange currents. It is
therefore realised that the concept of inert particles has only limited
value and that in the certain processes isobar degree of freedom should be 
considered. Thus a new picture of the nucleus emerged in which the nucleons
are not considered as being inert but having internal degrees of 
freedom(d.o.f). Such a configuration( called isobar configuration)
 is effectively described by allowing virtually excited isobar
to be present in the nucleus. These isobar configurations are dominated
by ${\bf{\triangle}}$ d.o.f[3,4]. Nucleonic interactions are clearly
influenced by the intrinsic excitation modes of the nucleon. At the 
center-of-mass excitation energy of roughly 300 Mev, pion-nucleon
reactions reveal a striking resonance behaviour in the $(J^{p},T) =
(\frac{3}{2}^{+},\frac{3}{2})$ channel. This resonance can be understood
as the formation of an excited state of the nucleon N(${1/2}^{+}$,1/2),
namely the $\triangle({3/2}^{+},3/2)$ and
is seen as strong resonance in $\Pi$N-Scattering.\\
The nucleon has many resonant states which are particularly  evident in pion-
nucleon reactions. These resonant states  can also be produced in the 
electromagnetic processes because most of them are coupled to the photon 
channel.  Some of such isobar states give rise to large enhancements in the total
absorption cross section. They are produced through different excitations
(i.e., M1, E1, and E2 for $\triangle$(1236), N(1520) and N(1680) respectively)
and decay mainly into the pion-nucleon channel. The study of such  transitions
is important from the point of view of the internal structure and should give
complementary informations with respect to what can be learned from the elastic
electron scattering. The quark contribution to the radiative transitions
between the hadrons has been investigated. the standard hypothesis is that a 
photon is emitted or absorbed by a single quark in the hadron and the
transition $\gamma X \rightarrow Y^{*}$ is thereby trigged. The most relevant
transition is  the one involving $\triangle$ ($\gamma \mbox{P}\rightarrow 
\triangle$) state. The process may be described in terms of a direct
$\gamma \mbox{N} \triangle$ coupling leading to an intermediate $\triangle$-
state, which subsequently decays into the $\Pi$-nucleon channel. There is
considerable interest in the helicity amplitudes for the above transition, 
since these determine the electricquadrupole $\mbox{E2}$ and magnetic
dipole $\mbox{M1}$ transition amplitude, the former depending on the \mbox{D}-
state percentage in the shell model quark wave function.\\
\section{$\triangle\rightarrow \mbox{N}\gamma$ Electromagnetic Transition}
The transition of ${3/2}^{+}$ to ${1/2}^{+}$ state takes place by interaction
with either the $\mbox{M}_{1}$ or $\mbox{E}_{2}$( the magnetic dipole or
electric quadrupole respectively)multipoles. It was noted by Becchi and Morpuro[5] that
in  constituent quark model the $\mbox{E}_{2}$ transition   is forbidden
in line with the data. This is essentially because the electric
quadrupole transition is proportional to the charge operator which
cannot cause transition between quark isospin 3/2 and 1/2 states
and hence the matrix element vanishes by orthogonality of the quark
wave functions. Also the
$\mbox{E}_{2}$ transition matrix element involves the spherical harmonics
$\mbox{Y}_{2}$ which cannot cause transition between $\mbox{L} = 0$ spectial 
wave functions as proton and $\triangle^{+}$ are both $\mbox{L} = 0$
in the quark model. The magnetic dipole transition involves the quark magnetic
moments and hence the spin-operator and this can lead to the transition
between 3/2 to 1/2 state[6]. The $\mbox{E}_{2}/\mbox{M}_{1}$ ratio thus vanishes
in the constituent quark model. But from the analysis of the experimental
data[7,8] this ratio is,
\begin{eqnarray}
\frac{\mbox{E}_{2}}{\mbox{M}_{1}} &=& -0.013\pm 0.005 \\
&=& -0.015\pm 0.002 
\end{eqnarray}
However subsequent introduction of the effect of quark-quark tensor and
spin-spin forces on the radiative decay of $\triangle$-isobar[9] leads to
a finite but rather very small $\mbox{E}_{2}/\mbox{M}_{1}$  ratio. In the
absence of the tensor force the nucleon and the isobar are both in the
state $S$ and their can be no $\mbox{E}_{2}$ transition.\\
To calculate the $\mbox{E}_{2}/\mbox{M}_{1}$ ratio(EMR) in the radiative
transition $\triangle$-isobar by making use of the helicity amplitudes,
we start with the current matrix element,
\begin{equation}
<\triangle_{p's'}\mid J_{\mu\nu}\mid N_{ps}> = \bar{U}_{\triangle}(p',s')
O_{\mu\nu}U_{N}(p,s)
\end{equation}
where $U_{N}$ is the nucleon spinor and $U_{\triangle}$ is the Rarita 
schwinger spinor which describes the $\triangle$-isobar. Taking both the
baryons on  mass shell, the Lorentz invariance and parity conservation
for the current yields the following form of the tensor;
\begin{equation}
O_{\mu\nu}= a_{1}[(g_{\mu\nu}q^{2}+p_{\mu}p_{\nu})]\gamma_{5}+a_{2}[g_{\mu\nu}(
M_{\triangle}^{2}-M_{N}^{2})+p_{\mu}P_{\nu}]\gamma_{5}+a_{3}[g_{\mu\nu}(
M_{\triangle}+M_{N})+p_{\mu}p_{\nu}]\gamma_{5},
\end{equation}
where,
\begin{eqnarray}
q &=&p'-p \nonumber\\   P &=& p'+p\nonumber
\end{eqnarray}
The functions $a_{i}$ are real and depend only on $q^{2}$. For excitation
through real photon $a_{1}$ does not contribute since $q^{2}=0$ and
$\varepsilon .q =0$, where $\varepsilon^{\nu}$ is the photon polarization
vector.\\
The form factors $a_{2}$ and $a_{3}$ are related to the functions $\mbox{G}_{i}$
of Pilkuhn[10] by;
\begin{eqnarray}
a_{2} & =& \mbox{G}_{2}/2(M_{\triangle}+M_{N})^{2} \nonumber\\
a_{3} &=& \mbox{G}_{1}/(M_{\triangle}+M_{N}) \nonumber
\end{eqnarray}
Using the explicit expressions for the spinors and keeping only the
lowest order terms in $P/M_{\triangle}$ and $P/M_{N}$, the current density
in non-relativistic limit becomes,
\begin{equation}
<\triangle_{p's'}\mid J\mid N_{ps}>={\Psi}_{\triangle}^{\dagger} \left[
a\sigma_{\triangle N}\bigg\{\sigma .(\frac{q}{\mu}-\frac{P}{\bar{\mu}}\bigg\}+
b\bigg\{(\sigma_{\triangle N} \times \sigma )\times (\frac{q}{\mu}-\frac{p}{
\bar{\mu}})\bigg\} \right]\Psi_{Ns},
\end{equation}
where,
\begin{eqnarray}
a &= & \frac{1}{2} \left[ a_{1}(M_{\triangle}-M_{N})+a_{2}(M_{\triangle}
+M_{N})-a_{3} \right](M_{\triangle}-M_{N}) \nonumber\\
b &= &\frac{1}{2}M_{N}a_{3} \nonumber\\
\frac{1}{\mu} &=& (\frac{1}{M_{N}} + \frac{1}{M_{\triangle}}) \nonumber\\
\frac{1}{\bar{\mu}} &= & (\frac{1}{M_{N}} - \frac{1}{M_{\triangle}})
\end{eqnarray}
The explicit  results are given in the appendix.\\
The magnetic dipole and electric quadrupole  form factors are given by,
\begin{eqnarray}
\mbox{G}_{M1}^{\triangle N} & = & \frac{2}{e(a-2b)} \nonumber\\
\mbox{G}_{E2}^{\triangle N} &=& \frac{4}{ea}
\end{eqnarray}
Putting Eq.(7) in Eq.(5) gives the following form of the current density;
\begin{equation}
J_{\triangle N} = e\left[-\frac{1}{2}\sqrt{5/3} \mbox{G}_{E2}^{\triangle N}
\bigg\{ \sigma_{\triangle N}^{[2]} \times \bigg(\frac{q^{[1]}}{2}-\frac{P^{[1]}}{2}\bigg)
\bigg\}^{[1]} +\frac{i}{4} \mbox{G}_{M2}^{\triangle N} \sigma_{\triangle N}
\times \bigg(\frac{q}{\mu}-\frac{P}{\bar{\mu}}\bigg) \right]
\end{equation}
where,
\begin{displaymath}
\sigma_{\triangle N}^{[2]} = \left[ \sigma_{\triangle N}^{[1]} \times 
\sigma^{[1]}\right]^{[2]}
\end{displaymath}
The helicity amplitudes are defined as,
\begin{equation}
A_{3/2} =  <\triangle \mid -e \overrightarrow{J}_{\triangle N}.
\overrightarrow{A} \mid N>
\end{equation}
 The relation between the multipole amplitudes and the helicity amplitudes
arise solely due to the resonance production $\mbox{N}\gamma \rightarrow
\triangle$.\\
Substituting Eq.(8) in Eq.(9), we get,
\begin{equation}
A_{3/2} = -\frac{\sqrt{\pi \omega}}{\sqrt{2M}}e^{2} \bigg(\mbox{G}_{M1}+
\mbox{G}_{E2} \bigg)
\end{equation}
similarly,
\begin{equation}
A_{1/2} = -\frac{\sqrt{\pi \omega}}{\sqrt{6M}} e^{2} \bigg(\mbox{G}_{M1}-
3\mbox{G}_{E2} \bigg)
\end{equation}
where,
\begin{displaymath}
\omega = M_{\triangle}-M
\end{displaymath}
is the energy of photon.\\
The ratio of helicity amplitudes is given by,
\begin{equation}
\frac{A_{3/2}}{A_{1/2}} = \sqrt{3}\frac{(1+\mbox{G}_{E2}/\mbox{G}_{M1})}
{(1-3\mbox{G}_{E2}/\mbox{G}_{M1})}.
\end{equation}
The helicity amplitudes, for the transition $\mbox{M1}$ and $\mbox{E2}$
nucleon resonance for proton excitation are[11]
\begin{eqnarray}
A_{3/2} &= & -179 \nonumber\\  A_{1/2} & =& -103 \nonumber
\end{eqnarray}
This yields an electric quadrupole to magnetic dipole transition ratio
for $\triangle \rightarrow \mbox{N} \gamma $ equal to,
\begin{displaymath}
\mbox{EMR} = E_{2}/M_{1} = -0.001
\end{displaymath}
\section{conclusion}
Although the importance of the $\triangle$-states in nuclear physics
has been widely studied, it has been only been possible to draw tentative
conclusions. The non-relativistic approach used for free $\triangle$
gives a good account of data but for intrinsically relativistic nuclear
many body problem the non-relativistic approximation seems to be less
important. The relativistic approach to nucleons and $\triangle$ in nuclear
matter may serve to help in clarifying this problem.
\appendix 
\section{$\triangle -\mbox{N}$ current}
The general expression for the current matrix element has the form;
\begin{equation}
<\triangle_{p's'} \mid J_{\nu} \mid \mbox{N}_{ps}>= 
\bar{\mbox{U}}_{\triangle}(p',s')O_{\mu\nu}(p',p)\mbox{U}_{N}(p,s)
\end{equation}
 where,
\begin{displaymath}
O_{\mu\nu}(p',p)= \bigg(\frac{c_{3}}{M}\gamma^{\nu}+\frac{c_{4}}{M^{2}}
p^{'\nu}+\frac{c_{5}}{M^{2}}p^{\nu}\bigg) \bigg(g_{\lambda\mu}g_{\rho\nu}
-g_{\lambda\rho}g_{\mu\nu}\bigg)q^{\rho}\gamma_{5}
\end{displaymath}
and $c_{i}$ are the Schiff and Tran functions.
Define,
\begin{displaymath}
a_{1}=\frac{c_{4}-c_{5}}{2M^{2}}, \hspace{1.5cm} a_{2}=\frac{c_{4}+c_{5}}{2M^{2}},
\hspace{1.5cm} a_{3}=\frac{c_{3}}{M}
\end{displaymath}
where $a_{i}$ are real and depend only on $q^{2}$.
The tensor becomes,
\begin{equation}
O_{\mu\nu}=\bigg(g_{\lambda\mu}a-q_{\lambda}b_{\mu}\bigg)
\end{equation}
with,
\begin{eqnarray}
a &= & (\frac{M_{\triangle}+M}{M})c_{3}+(\frac{c_{4}+c_{5}}{2M^{2}})(
M_{\triangle}^{2}-M^{2})+(\frac{c_{4}-c_{5}}{2M^{2}})q^{2}\nonumber\\
b_{\mu} &=& a_{3}\gamma_{\mu}+a_{2}p_{\mu}+a_{1}q_{\mu}\nonumber
\end{eqnarray}
Thus $\triangle - \mbox{N}$ current is given by;
\begin{equation}
J_{\mu}=\bar{U}(\overrightarrow{p'})\left[g_{\lambda\mu}a-\frac{q_{\lambda}}{M}b_{\mu}
\right]\gamma_{5}U(\overrightarrow{p})
\end{equation}
The leading term is,
\begin{equation}
J_{i}=c_{3}\bar{U}(\overrightarrow{p'})\left[g_{\lambda i}\frac{M_{\triangle}+M}{
M} - \gamma_{i}\frac{q_{\lambda}}{M}\right]\gamma_{5}U(\overrightarrow{p})
\end{equation}
Using the following form of the spinors;
\begin{eqnarray}
U(p) &=& \sqrt{ \frac{E+M}{2M}} \left( \begin{array}{c}
1 \\ \frac{\overrightarrow{\sigma}. \overrightarrow{p}}{E+M}
\end{array} \right) \Psi_{N} \nonumber\\
\bar{U}^{\lambda}(p') &= & \sqrt{ \frac{E+M_{\triangle}}{2M_{\triangle}}}
\left( \begin{array}{c}
1 \hspace{1cm} \frac{ \overrightarrow{\sigma}. \overrightarrow{p'}}{E+M_{\triangle}}
\end{array} \right) g^{\lambda} \Psi^{\dagger}_{\triangle} \gamma_{0},
\end{eqnarray}
where,
\begin{displaymath}
g^{\lambda}= L^{\lambda}_{\nu}(p') S^{\dagger \nu}
\end{displaymath}
and,
\begin{eqnarray}
g^{i} &= & -L^{i}_{j}S^{\dagger j} = \bigg(-M_{\triangle}S^{\dagger i}
+\frac{p.S^{\dagger}}{E+M_{\triangle}}p^{i} \bigg)/M_{\triangle} \nonumber\\
g^{0} &= & \frac{p_{i}.S^{\dagger}}{M_{\triangle}} \nonumber
\end{eqnarray}
the vector part of the current simplifies to,
\begin{equation}
J_{i}=\frac{a_{3}M}{2}\bigg( \overrightarrow{S} \times \overrightarrow{ \sigma}
\bigg) \times \bigg( \frac{ \overrightarrow{q}}{\mu} -\frac{ \overrightarrow{P}}{
\bar{\mu}}\bigg) - \frac{a_{3}}{4}\triangle M \overrightarrow{S}.\overrightarrow{\sigma}
\bigg( \frac{ \overrightarrow{q}}{\mu} - \frac{\overrightarrow{P}}{\bar{\mu}}
\bigg)
\end{equation}
The $a_{1}$ and $a_{2}$ terms are given by,
\begin{eqnarray}
X & \simeq & \frac{a_{1}}{4} \left[-S. \overrightarrow{\sigma}. \overrightarrow{A}
(M_{\triangle}-M)^{2} \right] \nonumber\\
Y & \simeq & \frac{a_{2}}{4}\left[-S.\overrightarrow{\sigma}.\overrightarrow{A}
(M_{\triangle}^{2}-M^{2}) \right]
\end{eqnarray}
where,
\begin{displaymath}
\overrightarrow{A} = \frac{1}{2}\bigg(\frac{\overrightarrow{P}}{\bar{\mu}} -
\frac{\overrightarrow{q}}{\mu}\bigg)
\end{displaymath}
Putting everything together gives the following form of $\triangle-\mbox{N}$
current;
\begin{eqnarray}
<\triangle \mid \overrightarrow{J} \mid N> & = &e\Psi^{\dagger}_{\triangle}
\left[a\overrightarrow{S}( \overrightarrow{\sigma}.\overrightarrow{A}) +
b(\overrightarrow{S} \times \overrightarrow{\sigma}) \times \overrightarrow{A}
\right]\Psi_{N} \nonumber\\
<\triangle \mid J_{0} \mid N> &=&e\Psi^{\dagger}_{\triangle}\left[
\frac{a}{M_{\triangle}-M}(\overrightarrow{S}.\overrightarrow{\sigma})
\overrightarrow{\sigma}.\overrightarrow{A} -\frac{b}{M_{\triangle}M_{N}}
\bigg( (\overrightarrow{S} \times \overrightarrow{\sigma}) \times
\overrightarrow{P} \bigg).\overrightarrow{q} \right]\Psi_{N}
\end{eqnarray}

\end{document}